\documentclass[twocolumn,showpacs,amsmath,amssymb]{revtex4}

\usepackage{graphicx}
\usepackage{dcolumn}
\usepackage{bm}
\usepackage{bbm}
\usepackage{amsfonts}

\newcommand{\slika}{3.2in}

\def\>{\rangle}
\def\<{\langle}
\def\ket#1{|#1\>}
\def\bra#1{\<#1|}
\def\braket#1#2{\< #1 | #2 \>}
\def\bracket#1#2#3{\< #1 | #2 | #3 \>}
\def\ave#1{\< #1\>}

\def\tr{{\,{\rm tr}}}

\def\im{{\,{\rm Im}\,}}

\newcommand{\op}[1]{\hat{#1}}

\begin{document}

\title{Evolution of entanglement under echo dynamics}
\author{Toma\v z Prosen$^1$, Thomas H. Seligman$^2$ and Marko \v Znidari\v c$^1$}
\affiliation{${}^1$ Physics Department, FMF, University of Ljubljana, Ljubljana, Slovenia\\
${}^2$ Centro de Ciencias F{\'\i}sicas, University of Mexico (UNAM), Cuernavaca,
Mexico}
\date{\today}
\begin{abstract}
Echo dynamics and fidelity are often used to discuss stability in quantum information 
processing and quantum chaos. Yet fidelity yields no information about entanglement, 
the characteristic property of quantum mechanics. We study the evolution of entanglement 
in echo dynamics. We find qualitatively different behavior between integrable and 
chaotic systems on one hand and between random and coherent initial states for 
integrable systems on the other. For the latter the evolution of entanglement 
is given by a classical time scale. Analytic results are illustrated numerically in a Jaynes Cummings model.

\end{abstract}

\maketitle

More than a century ago J. Loschmidt, in his discussions with L. Bolzmann illustrating
irreversibility in a gas, suggested to invert the velocity of each atom individually in order 
to revert to the initial situation. Recently Loschmidt echoes have been of great interest in the 
control of quantum information processing \cite{Nielsen}. As entanglement is the essential property of 
quantum mechanics, in the present paper we analyze the Loschmidt echo appearing in the 
entanglement of two quantum systems. As a measure of entanglement we use purity \cite{Zurek} and show 
analytically as well as numerically that for classically integrable systems the purity decays 
as $1-c_{\rm In}t^2$, whereas for a classically chaotic system the decay after the Zeno time 
scale is described by $1-c_{\rm ch}t$. For a coherent state the constant $c_{\rm In}$ is independent
of $\hbar$. For chaotic systems (or random initial states in integrable systems) this constant on 
the other hand is proportional to $1/{\hbar^2}$, thus defining quite different time scales.

Zurek \cite{Zurek} proposed to use the rate of decoherence as a characteristic of chaos in quantum mechanics and 
this is occasionally re-interpreted in terms of entanglement \cite{Nemes} between different parts of a 
closed quantum system. Such studies were limited to forward time evolution. Yet the  insensitivity of 
quantum mechanics to small 
changes in initial conditions has been a basic difficulty in the introduction of the concepts of 
chaos to this field, 
and the idea to use sensitivity to perturbations in the time evolution \cite{Peres} has emerged 
recently as one 
of the tools of choice to overcome this difficulty \cite{Jalabert, Ktop}. Specifically all these 
authors use {\em fidelity, i.e.} the correlation function between a quantum state evolving 
under the action of two Hamiltonians differing by a perturbation, which is equivalent to the 
autocorrelation function under echo dynamics.
Since fidelity measures irreversibility of a full quantum state under the echo,
it is also desirable to undertand irreversibility of a less restrictive 
quantity like entanglement. A recent study in spin 
chains \cite{Spinecho}, more related to quantum computing, revealed no qualitative difference between 
fidelity and the evolution of entanglement under echo dynamics as measured conveniently by purity there denoted as {\em
purity-fidelity}.
This system allows no classical analogue, and by consequence no coherent states. Yet these
we shall show to be essential to recover results analogue to the ones of Zurek and Nemes
\cite{Zurek,Nemes}. Based on the idea that a partial trace simulates decoherence our results lead to the 
conjecture that Zureks result holds exclusively for coherent states in the central system, i.e.
that we may not expect faster decoherence for chaotic central systems than for integrable
ones, if we use random states that are more relevant to quantum information.

To implement this idea we have to consider systems with at least two degrees of freedom
to allow for entanglement and a well defined classical limit $\hbar \to 0$. 
The (unperturbed) Hamiltonian will contain a parameter
permitting a transition from order to chaos, and will typically couple the two degrees of
freedom. A perturbation is then defined to obtain a second similar Hamiltonian.
We give general results for entanglement under echo-dynamics starting from an initial 
product (dis-entangled) state.
To illustrate our results we use the Jaynes Cummings (JC) model \cite{Jaynes&Cummings}.
The usual co-rotating (integrable) version of this model
has great practical importance in atomic physics and illustrates the $\hbar$ independent evolution
of entanglement for coherent states.
For the arguments involving the chaotic dynamics we include counter-rotating terms \cite{Nemes, Debergh}
to construct a toy model that allows for chaos.
Even this model may not be entirely unrealistic for atoms in a Paul trap in a driven field, as standard 
papers seek conditions where this term is small\cite{Cirac}.

For general considerations and analytic calculations 
techniques of linear response developed originally for the evaluation of fidelity \cite{Ktop} 
are extended to calculate purity-fidelity in terms of time correlation functions 
of the perturbation. In the case of coherent states we could carry the evaluation of linear response
one step further using it in a semi-classical framework that relates the decay rates directly
to the stability matrix of the orbit along which the packet evolves.
     
\par
We consider the unitary time evolution
given by the echo operator $M_\delta(t)=U^\dagger_\delta(t)U(t)$. Here $U(t)$ 
is generated by some unperturbed Hamiltonian $H$ as $U(t)=e^{-iHt/\hbar}$
and similarly $U_\delta(t)=e^{-i(H+\delta V)t/\hbar}$, where $V$ is the perturbation with strength $\delta$.
It is useful to rewrite the echo operator as time-ordered 
product \cite{Ktop} in the interaction picture
\begin{equation}
M_\delta(t)=\op{\cal T}\exp\left({\rm i}\Sigma(t)\delta/\hbar\right),
\label{eq:Mt}
\end{equation}
where $\Sigma(t):=\int_0^t{V(\tau)d\tau}$ with $V(t):=U^\dagger(t) V U(t)$.
This operator shall act on a composite system with the Hilbert space
${\cal H}={\cal H}_1\otimes {\cal H}_2$,
consisting of two factors with dimensions $N_1$ and $N_2$, 
which we may look upon as a ``central system'' and an ``environment''.
We are interested only in information about the subsystem $1$
which is contained in a {\em reduced density matrix} 
$\rho_1(t):={\rm tr_2}\rho(t)$, $\rho(t)=\ket{\psi(t)}\bra{\psi(t)}$.
We shall study the purity-fidelity \cite{Spinecho} $F_P(t):=\tr_1[\rho_1(t)]^2$ as a
measure of factorizability of a joint state $\ket{\psi(t)}=M_\delta(t)\ket{\psi(0)}$.
 We choose this quantity rather than some entropy because 
of its simple analytic dependence on $\rho_1(t)$.
Here the partial traces with indices 1 and 2 are taken 
in the corresponding factor spaces and the reduced density matrix 
is acting on the first factor space. We always 
assume that we start with a factorized state at $t=0$, {\it i.e.} $F_P(0)=1$.
For comparison we shall also use the fidelity $|F(t)|^2 = |\bra{\psi(0)}
M_{\delta}(t)\ket{\psi(0)}|^2$.

Expanding the echo operator (\ref{eq:Mt}) in $\delta$, we get \cite{Ktop}
\begin{eqnarray}
|F(t)|^2 &=& 1- \delta^2 \hbar^{-2} C(t) + \cdots, \label{eq:F2nd} \\
C(t)&:=& \ave{\Sigma^2(t)} - \ave{\Sigma(t)}^2.
\nonumber
\end{eqnarray}       
Here $\ave{.}$ denotes an expectation in the product initial state $\ket{\psi(0)}=\ket{1,1}$ with the 
abbreviation $\ket{i,\nu }:=\ket{i}_1 \otimes \ket{\nu}_2$.
The same techniques yield for purity-fidelity
\begin{eqnarray}
F_{\rm P}(t) &=& 1-2 \delta^2 \hbar^{-2} \left\{C(t) - D(t)\right\} +\cdots, \label{eq:Fp2nd} \\
D(t) &:=& 
\sum_{\nu \neq 1} |\bracket{1,\nu}{\Sigma(t)}{1,1}|^2 + \sum_{i \neq 1}{|\bracket{i,1}{\Sigma(t)}{1,1}|^2} 
\nonumber
\end{eqnarray}
For both series to converge it is sufficient to use a {\em bounded} perturbation operator $V$,
but we expect the linear response formula to be a good approximation for a much wider class of
perturbations. 
The somewhat unusual correlation function $D(t)$ contains only off-diagonal matrix elements 
of the operator $\Sigma(t)$ and determines the difference between $F_P(t)$ and $|F(t)|^4$. 
From expansions (\ref{eq:F2nd},\ref{eq:Fp2nd}) we can see that the decay is determined by time
correlation functions of the perturbation. 
The stronger the decay of correlation functions $\bra{\psi}V(t)V(t')\ket{\psi}$ 
as $|t-t'|$ grows, the slower is the increase of $C(t)$ and the slower is the decay of $F(t)$ and 
$F_{\rm P}(t)$.
\par
We limit our discussion to systems which have a classical limit.
For such systems chaos typically implies decay of the time correlation functions of the perturbation observable
({\it i.e.} mixing), while regular motion implies non-ergodic behavior.
Fidelity decay for both situations is discussed in Ref.~\cite{Ktop}.
Under rather general assumptions one finds {\em exponential} decay for {\em chaotic} dynamics
\begin{equation}
|F(t)|^2=\exp{(-t/\tau_{\rm em})}, \qquad \tau_{\rm em}= (2 \sigma)^{-1} \hbar^{2} \delta^{-2},
\label{eq:Fexp}
\end{equation}
where a {\em diffusion coefficient} $\sigma:=\lim_{t \to \infty}C(t)/(2t)$ is independent of the 
initial state $\ket{\psi(0)}$ [for sufficiently long times, typically $t \gg \log 1/\hbar$]. 
In classically {\em regular} situation the fidelity exhibits a quadratic decay in the 
leading order in $\delta$ even for long times, since
$C(t)\to \bar{c}t^2$, where $\bar{c}$ depends on the structure of the initial state.
For a coherent initial state we find a {\em Gaussian} decay of fidelity
\begin{equation}
|F(t)|^2=\exp{(-(t/\tau_{\rm ne})^2)}, \qquad
\tau_{\rm ne}=\bar{c}^{-1/2} \hbar \delta^{-1}
\label{eq:Fgauss}
\end{equation}
with $\bar{c} \propto \hbar$ \cite{Ktop}.
It is worth to stress that in the regime of linear response (small $\delta$) formulae (\ref{eq:Fexp},\ref{eq:Fgauss})
agree with Eqs. (\ref{eq:F2nd},\ref{eq:Fp2nd}) from which the time scales $\tau_{\rm em},\tau_{\rm ne}$ are obtained.
\par
As for purity-fidelity of chaotic systems, one may argue that $\Sigma(t)$ should look like
a {\em random matrix} so the term $D(t)$ should be small compared
to $C(t)$ in (\ref{eq:Fp2nd}), namely $D(t)/C(t) \sim 1/N_1 + 1/N_2$ because of the 
smaller number of terms involved in the sums.
Thus, if both dimensions $N_{1,2}$ grow as $\hbar\to 0$ one expects in the asymptotic regime
that $F_{\rm P}(t) = |F(t)|^4 = \exp(-2t/\tau_{\rm em})$, and does not significantly depend 
on the initial state. In particular this also holds for coherent initial states.
Using similar arguments for regular dynamics but a random initial state one again sees that
$F_{\rm P}(t)$ follows $|F(t)|^4$ closely \cite{Spinecho}. 

Yet for {\em coherent states} and regular classical dynamics
this is {\em not} the case because the term $D(t)$ 
is not negligible. We show that the difference $C(t)-D(t)$ cancels in the leading order in $\hbar$, 
i.e. $C(t)-D(t) \sim \hbar^2$, meaning that $F_{\rm P}(t)$ as 
compared to $|F(t)|^2$ decays on a qualitatively longer, $\hbar$-independent time scale 
$\tau^{\rm P}_{\rm ne} = K/\delta \sim \hbar^{-1/2}\tau_{\rm ne}$. This will be the main result of 
the present paper. In order to establish this we consider the evolution of a Gaussian wave-packet 
along a stable orbit $\vec{z}_t=(\vec{x}_t,\vec{p}_t)$ as 
$\braket{\vec{x}}{\psi(t)} = 
C\exp(\frac{i}{\hbar}[(\vec{x}-\vec{x}_t)\cdot A_t(\vec{x}-\vec{x}_t) + \vec{p}_t\cdot\vec{x}])$,
where the block-form of the complex $d\times d$ matrix
\begin{equation}
A_t = 
\begin{pmatrix}
A_{11} & A_{12}\cr
A_{21} & A_{22}\cr 
\end{pmatrix}
\end{equation}
corresponds to obvious division of $d=d_1 + d_2$ dimensional
configuration space into $d_1$ and $d_2$ dimensional parts.
$A_t$ is a ratio of two pieces of a classical monodromy matrix \cite{Heller} so it is 
$\hbar$-independent. The purity of a reduced wave-packet $\rho(x_1,x'_1)=\int d x_2 
\braket{x_1,x_2}{\psi(t)}\braket{\psi(t)}{x'_1,x_2}$
is $F_P=\int d x_1 d x'_1 |\rho(x_1,x'_1)|^2 = 1$ if $A_{12}=A_{21}=0$ while in general we find
$\hbar-$independent expression
$$
F_P = (\det \im A)
\left|\begin{matrix}
\im A_{11} & \frac{\rm i}{2} A_{12}^*  & 0 & -\frac{\rm i}{2} A_{12} \cr
         \frac{\rm i}{2} A_{21}^*  & \im A_{22} & -\frac{\rm i}{2} A_{21} & 0 \cr
         0 & -\frac{\rm i}{2} A_{12} & \im A_{11} & \frac{\rm i}{2} A_{12} \cr
         -\frac{\rm i}{2} A_{21} & 0 & \frac{\rm i}{2} A_{21}^*  & \im A_{22}\cr 
\end{matrix}\right|^{-1/2}.
$$
For classical echo dynamics, the covariance matrix $A=A_t$ is given by a linear stability analysis as 
$A = A_0 + t\delta B$ for some matrix $B$, where 
$A_{0,12}=A_{0,21}=0$. Then purity-fidelity is $\hbar$-independent and can be evaluated in the leading orders as 
$F_{\rm P}(t) = 1 - (t \delta/K)^2 + \ldots$. 

We thus reach the following interesting conclusion: Both 
fidelity and purity-fidelity decay quadratically in integrable situations, while they decay linearly 
in chaotic ones, once we are beyond the Zeno time scale.
Yet there is a very relevant difference in time scales themselves, if we discuss the purity of coherent rather 
than random states. 
For integrable systems purity-fidelity decays on an $\hbar$-independent scale. This leads to situations with very 
stable purity-fidelity, 
while the same perturbation generates decay of the fidelity of the coherent state as well as 
the decay of the purity-fidelity of a random state on  much shorter time scales, dictated by the value of $\hbar$. 
Note though that for sufficiently small perturbations at fixed $\hbar$ the quadratic 
decline of purity-fidelity always prevails.

To illustrate these results we use the JC Hamiltonian including  co- {\em and } counter-rotating 
terms for the chaotic case as
\begin{equation}
H= \hbar \omega a^\dagger a+ \hbar \epsilon J_{\rm z} + \frac{\hbar}{\sqrt{2J}} (G a J_{+} + G' a J_{-} + {\rm h.c.})  
\label{eq:JC}
\end{equation} 
with standard boson operators $a,a^\dagger$, $[a,a^\dagger]=1$, and 
standard SU(2) generators $J_{\pm},J_{\rm z}$.
We choose $\hbar=1/J$ ensuring that the classical limit is reached for $J \to \infty$ while the
angular momentum $\hbar J=1$ is fixed.
If either $G=0$ or $G'=0$ the model is integrable with an additional 
invariant being the difference or the sum of quanta for the spin and the oscillator.
In all calculations we used coherent 
initial states for the product system, i.e. direct product of coherent states
of the oscillator,
$\ket{\alpha}_{2}=e^{\alpha a^\dagger-\alpha^* a}\ket{0}_2$,
and of the spin [SU(2)],
$\ket{\theta,\phi}_1=(1+\tau \tau^*)^{-J} 
\exp{(\tau J_{-})}\ket{J,J}$ with $\tau=e^{{\rm i}\phi} \tan{(\theta/2)}$ \cite{Agarwal81}.

For our numerics we fix $J=4$ and choose initial position of SU(2) coherent state at $(\theta,\phi)=(1,1)$ and 
for the oscillator at $\alpha=1.15$. The parameters in JC Hamiltonian are: (a) in chaotic regime 
$\omega=\epsilon=0.3$ and $G=G'=1$, (b) in integrable regime $\omega=\epsilon=0.3$ and $G=1,G'=0$. 
The corresponding classical Poincar\' e section shows a single practically ergodic component in
the chaotic case (a) [at energy $E=1.0$ determined from the initial condition], whereas integrable case (b) 
[$E=0.63$] shows a generic family of invariant tori. 
The perturbation is realized by varying the parameter $\epsilon$ in JC Hamiltonian 
(\ref{eq:JC}), also known as dephasing, so the (bounded) perturbation generator is $V=\hbar J_{\rm z}$. 

\begin{figure}
\centerline{\includegraphics[width=\slika]{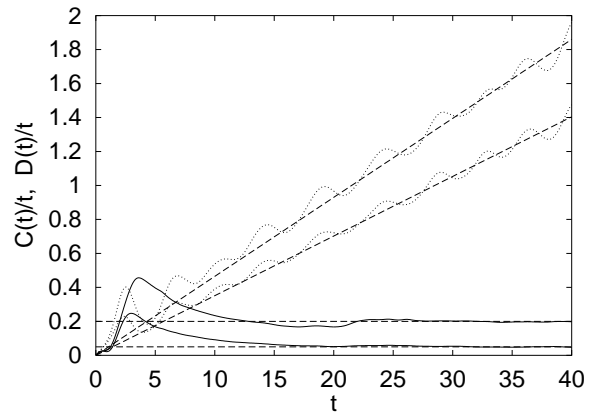}}
\caption{Correlation integrals $C(t)/t$ and $D(t)/t$ for regular (top two dotted curves) and 
chaotic regime (lower two solid curves). In both cases the upper curve is for $C(t)/t$ and the lower for $D(t)/t$. 
The horizontal
dashed lines indicate $2\sigma=0.20$ (upper) and $0.20/4$ (lower),
whereas the increasing ones have the slopes $\bar{c}=0.046$ (upper) and $(1-0.98/4)\bar{c}$ (lower).}
\label{fig:corr}
\end{figure}

We now show numerical results obtained by diagonalization in truncated Hilbert spaces.
Stability of the calculation with respect to truncation was checked.
Fig.~\ref{fig:corr} presents the correlation integrals $C(t)/t$ and 
$D(t)/t$ for chaotic and regular regimes. For chaotic dynamics 
(case a) the correlation integral converges after $t \approx 10$ to a well defined 
diffusion coefficient $\sigma=0.10$ with the D-term being of order $1/N_1+1/N_2\approx 1/4$. 
For regular dynamics (case b) and $t>10$ the correlation integral grows as $t^2$ due to a 
non-vanishing plateau $\bar{c}=0.046$ in the correlation function. In this case the 
difference $C(t)-D(t)$ is approximately $C(t)/J \propto \hbar^2$, which has been checked numerically
also for larger $J\le 24$, confirming $\hbar$-independent decay of $F_{\rm P}(t)$.
The oscillations in these functions are not accounted for by the present theory,
and are probably particular but interesting properties of the model. Whether they relate to 
oscillations seen in \cite{Nemes} is an open question.

\begin{figure}
\centerline{\includegraphics[width=\slika]{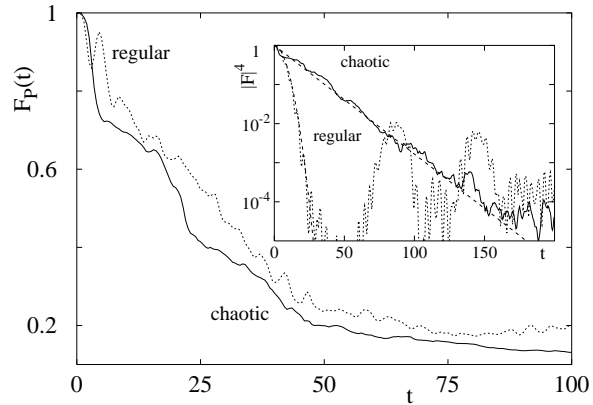}}
\caption{Purity-Fidelity $F_{\rm P}(t)$ (main figure) and squared fidelity $|F(t)|^4$ (inset) 
in chaotic regime (solid curves) and in integrable regime (dotted curves), for $\delta=0.1$.
The dashed lines indicate the linear and quadratic approximation respectively.
Note the differences in vertical scales.
}
\label{fig:purf}
\end{figure}

We first report a calculation with a strong perturbation  $\delta=0.1$, that rapidly exceeds  the 
realm of validity of linear response, in Fig.~\ref{fig:purf} where the main figure gives the 
purity-fidelity and the inset the fidelity.
For the fidelity decay (inset) we find excellent agreement with the
exponential decay (\ref{eq:Fexp}) in a chaotic 
regime and a faster Gaussian decay in a regular regime (\ref{eq:Fgauss}), where the 
decay rates are fixed as above.
However, for purity-fidelity we find 
already at $t \approx 20$, that the decay starts to be influenced by the saturation value of 
$F_{\rm P}(t \to \infty) \approx 1/(2J+1)$. Therefore purity-fidelity is
higher for the integrable case than for the chaotic one  not only at short times, as expected, but
even at large times. This is relevant, because we shall next choose a weak perturbation $\delta=0.005$
to avoid this problem. 
\begin{figure}
\centerline{\includegraphics[width=\slika]{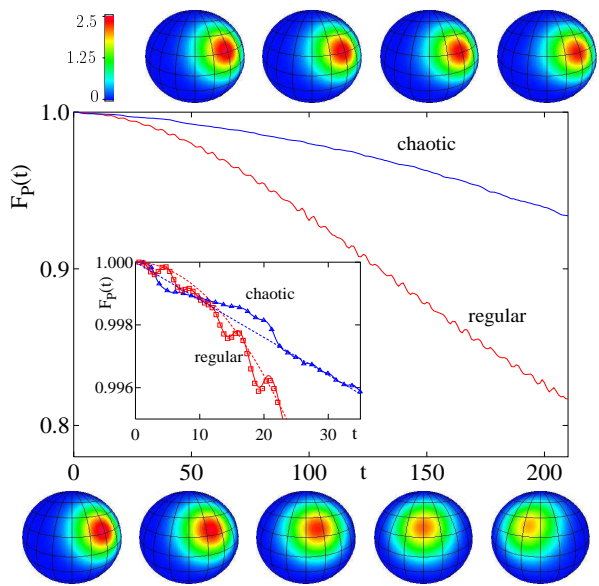}}
\caption{(Color)~Echo dynamics for weak coupling: $\delta=0.005$.
 Square of the Wigner function for chaotic dynamics (top diagrams) and integrable dynamics as a function of 
time (bottom diagrams) at times corresponding to the axis.
Color code: top left.
Purity-fidelity is shown in the frame on the same time scale and for short times in the inset. 
Red curves give the integrable and blue curves the chaotic evolution. Full curves show the complete numerics, 
symbols the evaluation starting from the numerical correlation functions of Fig.~\ref{fig:corr} and dashed curves 
the linear or quadratic approximation.}
\label{fig:wig}
\end{figure}
We expect and find the crossover after a fairly short time.
This calculation allows comparison with theory as well as an illustration 
of the evolution of the square of the Wigner function , corresponding to the reduced density matrix
$\rho_1(t)$ for the angular momentum  states on the sphere 
using the definition of \cite{Agarwal81}. 
Near the top and bottom of Fig.~\ref{fig:wig} 
we see this evolution for the chaotic and the integrable Hamiltonian 
respectively. 
In the center of the figure we plot the purity-fidelity on the same time scale as the Wigner 
functions in the main frame and an amplification of short times in the inset.
We observe detailed agreement of numerics with results obtained from the numerical values of
the correlation integrals (\ref{eq:F2nd},\ref{eq:Fp2nd}) reproducing the 
oscillatory structure. From the same correlation integrals we obtained the coefficients 
for the linear and quadratic decay, which  agree well if we discard the oscillations. 
We see a crossing of the two curves at $t=t^*_{\rm P}\approx 12$ 
for $F_{\rm P}$.
This times are larger than the Zeno time ($\approx 1$) and indicate 
the competition of the decay rate and the decay shape as expected for a non-small value of $\hbar=1/4$.
It is important to remember that the integral over the square of the Wigner function gives the purity
and therefore the fading of the picture will be indicative of the purity decay. On the other hand the 
movement of the center is an indication of the rapid decay of fidelity 
(not shown in the figure).

In this letter we study the linearized behavior of the evolution of entanglement under echo dynamics for time-scales large compared to those of the 
quantum Zeno effect, but 
sufficiently short for the expansion to be valid.
Similarly to the behavior of fidelity the decay of purity-fidelity is 
typically quadratic for non-mixing systems, and linear for mixing ones, the first situation arising for 
integrable systems and the second for chaotic ones. 
An interesting particular, but relevant case appears if we consider coherent initial 
states and integrable classical dynamics. In this case we have shown that purity-fidelity, 
still having a quadratic decay, can be computed classically in the leading order which is $\hbar$-independent,
so the time scale for purity-fidelity decay of a coherent state is longer by a factor proportional to $\hbar^{-1/2}$ than 
the corresponding one for a random state. Coherent states
in integrable systems are thus particularly long lived for semi-classical echo situations.
On the other hand, for chaotic classical dynamics and coherent initial states we find that purity-fidelity is the same as for random
states, and its decay will be slower than for either random or coherent initial states and integrable dynamics provided that time is sufficiently long or
perturbation $\delta$ is sufficiently small.
\par
We are grateful to W. Schleich for his extensive discussion of our manuscript.
Financial support by the Ministry of Education, Science and Sports of Slovenia  
and from projects IN-112200, DGAPA UNAM, Mexico,  25192-E CONACYT Mexico and
DAAD19-02-1-0086, ARO United States is gratefully acknowledged. 
TP and M\v Z thank CIC,  where this work was completed, for its hospitality.


\begin{references}

\bibitem{Nielsen} M.~A.~Nielsen and I.~L.~Chuang {\em Quantum Computation and Quantum information}
(Cambridge University Press, Cambridge 2000).

\bibitem{Zurek} W.~H.~Zurek, Physics Today {\bf 44}(10), 36 (1991);
W.~H.~Zurek and J.~P.~Paz, Physica {\bf 83D}, 300 (1995).

\bibitem{Nemes} K.~Furuya {\em et al.}
Phys.~Rev.~Lett. {\bf 80}, 5524 (1998);
R.~M.~Angelo {\em et al.} Phys.~Rev.~A {\bf 64}, 043801 (2001).

\bibitem{Peres} A.~Peres, Phys.~Rev.~A {\bf 30}, 1610 (1984).

\bibitem{Jalabert}
H.~M.~Pastawski {\em et al.} Phys.~Rev.~Lett. {\bf 75}, 4310 (1995);
P.~R.~Levstein {\em et al.} J.~Chem.~Phys. {\bf 108}, 2718 (1998);
R.~A.~Jalabert and H.~M.~Pastawski, Phys.~Rev.~Lett.~ {\bf 86}, 2490 (2001);
Ph.~Jacquod {\em et al.} Phys.~Rev.~E {\bf 64}, 
055203(R) (2001); N.~R.~Cerruti and S.~Tomsovic, Phys.~Rev.~Lett. {\bf 88}, 054103 (2002);
F.~M.~Cucchietti {\em et al.} Phys.~Rev.~E {\bf 65} 046209 (2002); D.~A.~Wisniacki and D.~Cohen, 
 Phys.~Rev.~E {\bf 66} 046209 (2002); T.~Kottos and D.~Cohen, Europhys.~Lett.~ {\bf 61} 431 (2003).

\bibitem{Ktop} T.~Prosen and M.~\v Znidari\v c, J.~Phys.~A {\bf 35}, 1455 (2002);
T.~Prosen, Phys.~Rev.~E {\bf 65}, 036208 (2002).

\bibitem{Spinecho} T.~Prosen and T.~H.~Seligman, J.~Phys.~A {\bf 35}, 4707 (2002).

\bibitem{Jaynes&Cummings} E.~T.~Jaynes and F.~W.~Cummings, Proc. IEEE {\bf 51}, 89 (1963); M.~Tavis and 
F.~W.~Cummings, Phys.~Rev. {\bf 170}, 379 (1968).





\bibitem{Debergh} See, {\it e.g.}, N.~Debergh and A.~B.~Klimov, Int.~J.~Mod.~Phys.~A {\bf 16}, 4057 (2001).

\bibitem{Cirac} See, {\it e.g.}, J.~I.~Cirac et al., Phys.~Rev.~Lett. {\bf 70}, 762 (1993)

\bibitem{Heller} E.~J.~Heller, in {\em Chaos and Quantum Physics}, ed. A. Voros {\em et al}
(North-Holland, Amsterdam 1991).

\bibitem{Agarwal81} G.~S.~Agarwal, Phys.~Rev.~A {\bf 24}, 2889 (1981).

\end{references}
\end{document}